\newcommand{\eq}[1]{Eq.~(\ref{eq:#1})}
\newcommand{\fig}[1]{Fig.~\ref{Fig:#1}}
\newcommand{\sect}[1]{Section~\ref{sect:#1}}

\documentclass[preprint,aps,prl,showpacs,preprintnumbers,superscriptaddress
,amsmath,amssymb]{revtex4}
\usepackage{graphicx}
\usepackage{dcolumn}
\usepackage{bm}
\usepackage[usenames]{color}
\usepackage{slantsc}
\usepackage{dsfont}
\newcolumntype{d}{D{.}{.}{2.2}}
\usepackage{siunitx}

\begin{document}

\title{Monte Carlo calculations for Fermi gases in the
unitary limit with a zero-range
interaction}

\author{Renato Pessoa}
\email{rpessoa@ufg.br}
\address{Instituto de F\'{\i}sica, Caixa Postal 131 \\
Universidade Federal de Goi\'as (UFG) \\
74001-970, Goi\^ania, GO, Brazil}
\address{Department of Physics and Astronomy, \\
Arizona State University \\
85287, Tempe, Arizona, USA}
\author{S. A. Vitiello}
\email{vitiello@ifi.unicamp.br}
\address{Instituto de F\'{\i}sica Gleb Wataghin, Caixa Postal 6165 \\
Universidade Estadual de Campinas - UNICAMP \\
13083-970, Campinas, SP, Brazil}
\author{K.~E.~Schmidt}
\address{Department of Physics and Astronomy, \\
Arizona State University \\
85287, Tempe, Arizona, USA}

\date{\today}


\begin{abstract} An ultracold Fermi gas with a zero-range
attractive potential in the unitary limit is investigated using variational
and diffusion Monte Carlo methods. Previous calculations have used a
finite range interactions and extrapolate the results to zero-range.
Here we extend the quantum Monte Carlo method to directly use a
zero-range interaction without extrapolation.
We employ a trial wave function with the correct
boundary conditions, and modify the sampling procedures to handle
the zero-range interaction.
The results are reliable and have low variance.
\end{abstract}

\maketitle

\section{Introduction}
\label{intro}

One of the main characteristics of ultracold atomic
Fermi gases is that the range of the
interatomic potential is much smaller than the average distance between
the particles.
This eliminates the range of the interaction as one of the
parameters of the system
and makes the system have simple scaling and universality.
Only the \textit{s}-wave states are important for the two
body scattering. Despite the very short range of the interatomic interaction,
it can have a large effect when the scattering length is large.
The unitary
limit
is characterized by
a scattering length $a$ satisfying $a k_F\rightarrow - \infty$, where $k_F$ is
the Fermi momentum,
and corresponds to the BEC-BCS crossover region.
This strongly correlated regime is governed by universal
relations associated with the large value of the scattering
length~\cite{bra10,ket08,gan11}. When the potential range
is negligible the details of the interactions the only length scales
are given by the scattering length and $k_F^{-1}$ which is determined
by the particle density.
For infinite scattering length,
the ground state energy per particle of the system
is proportional to the energy of the free Fermi gas, $E/N = \xi E_{FG}$, where
$\xi$ is the Bertsch parameter and $E_{FG} = \frac{3}{5}E_{F}$ is the
energy per particle for a noninteracting Fermi gas and $E_{F}$ is the
Fermi energy.

Quantum Monte Carlo (QMC) methods~\cite{gio08} are a set of tools
that can be used to study strongly correlated systems. The first successful
study of ultracold Fermi gases using QMC was made by Carlson \textit{et
al.}~\cite{car03} more than ten years ago. QMC has been
used to calculate various
properties of ultracold gases such as the pairing gaps
and energies for different values of scattering
length~\cite{cha04,gez08,car08},  the momentum distribution~\cite{ast05,gan11},
pair distribution functions~\cite{ast04,cha05} and condensate
fraction~\cite{ast05}. 
These previous quantum Monte Carlo calculations
have used a parameterized short, but finite, effective range model potential.
A series of calculations are done with different values of the effective
range are then extrapolated to zero-range. Care needs to be used, since
standard variational or diffusion Monte Carlo methods give a growing
variance as the range of the potential is decreased.

In this paper we show how to perform QMC calculations using
a zero-range interaction. Since with Monte Carlo sampling, the
probability of sampling two particles exactly at contact is zero,
we must construct trial wave functions that have the correct behavior
at contact. This imposes a boundary condition on the 
trial wave function. By enforcing this boundary condition
on our trial wave function,
we can eliminate the extrapolation to zero-range in variational Monte
Carlo caluclations. For diffusion Monte Carlo calculations, the usual
Trotter breakup does not go through since the commutator terms diverge.
Instead, we must solve for the two-body propagator exactly with the zero-range
potential. This calculation is easily done analytically, and we have
developed methods for sampling this propagator.

Calculations with a finite effective range potential, and standard
Jastrow-BCS trial wave functions show a diverging
variance when the effective range is small. This behavior requires
that we also modify the sampling
for both variational and diffusion Monte Carlo to reduce this variance,
with our choices of the trial wave function. The correct wave function
would not have these divergences, but we have not been able to write
a trial wave function for the zero-range potential that is computationally
efficient and removes these variances. The energy expectation value
is well behaved, so by modifying our sampling procedure, we are able
to cancel the divergent pieces
and obtain reliable results even in the zero-range limit.

The paper is organized in the following way. In \sect{MC} the QMC
calculations are discussed for the zero-range potential system in the
unitary limit. In \sect{details} we present the computational details
used to perform the calculations. The results are presented and discussed
in \sect{results} and, in \sect{summ} we give a summary and conclusions.

\section{Monte Carlo methods}
\label{sect:MC}

The Hamiltonian for the unpolarized unitary Fermi gas with $N$ particles
of mass $m$ is
\begin{equation}
\label{eq:Hamiltonian}
H =\sum_i^{N/2} \frac{p_i^2}{2m}
+\sum_{i'}^{N/2} \frac{p_{i'}^2}{2m}
+ \sum_{i,i'} v(|\vec r_i-\vec r_{i'}|)
\end{equation}
where the primed index denotes the spin-down particles and the unprimed index
the spin-up particles.
The potential $v(r)$ is
attractive and has range that goes to zero keeping the scattering
length fixed.

The trial wave function $\Psi_T$ that we use is of the Jastrow-BCS form
\begin{eqnarray}
\label{eq:PsiT}
\Psi_T(R) &=& F(R)\Phi_{JBCS}(R)
\nonumber\\
F(R) &=&
 \prod_{i,i'} f_{\uparrow\downarrow}(r_{i,i'})
\prod_{i<j} f_{\uparrow\uparrow}(r_{i,j})
\prod_{i'<j'} f_{\uparrow\uparrow}(r_{i',j'}) 
\nonumber\\
\Phi_{JBCS}(R) &=&
\mathcal{A}\left [
\phi(\vec r_1-\vec r_{1'})
\phi(\vec r_2-\vec r_{2'})
...
\phi(\vec r_{N/2}-\vec r_{N/2'}) \right ]
\end{eqnarray}
where $R \equiv \{\mathbf{r}_1,\mathbf{r}_2, \ldots,
\mathbf{r}_1',\mathbf{r}_2', \ldots \}$ stands for the positions of
all the particles.
The symmetrical Jastrow factors $f_{\uparrow\uparrow}(r)$
correlate like spin pairs,  and $f_{\uparrow\downarrow}(r)$ correlate
unlike spin pairs.
$\mathcal{A}$ is an antisymmetrizer
operator which antisymmetrizes like spin particles, and $\phi(\vec r)$ is
the Cooper pair wave function that gives BCS pairing. Since the
purpose of this paper is to demonstrate the algorithm, we take 
\begin{equation}
\phi(\vec r) = \sum_{\vec k,k< k_{FB}} \cos(\vec k \cdot \vec r)
\end{equation}
where $k_{FB}$ is the Fermi wave vector for our finite periodic system.
In this case, the BCS form reduces to a product of two Slater determinants
representing the ground state of the noninteracting system.
Fixed node
calculations for this trial function have been done for finite
range interactions with extrapolations to zero-range, which will allow
comparisons to our method.

The BCS function defines the nodal surface structure of the system. The
nodes are fixed and we deal with the sign problem using
the fixed-node approximation i.e. forbidding the system
to cross the nodal surface.
While we use the Jastrow-Slater limit here, we plan to optimize our
results in the future
with the more general Jastrow-BCS wave function that has been widely
used in the literature~\cite{car03,cha04,ast05,gan11}.

The trial wave function typically has adjustable parameters to obtain
the upper bound limit for the ground state energy of the system in
accordance with the variational principle. The quality of the results,
obtained through the variational Monte Carlo method (VMC), usually depends
on the quality of the trial wave function employed in the calculations.
These results can be improved by introducing parameters into the
trial wave function
based on physical
insight.

To use the zero-range potential, we can
look at the limit of the behavior of
the system when two unlike spin particles when the potential is
of a very small, but finite range. In that case, when the pair is
within the range of the potential, the value of the potential must
go to negative infinity as the range goes to zero. This potential
then completely dominates the Schr\"odinger equation, and leads to
the boundary condition
when
$r_{i,i'}\rightarrow 0$
\begin{equation}
\Psi(R) = A \left( \frac{1}{r_{i,i'}} - \frac{1}{a} \right),
\end{equation}
as
derived by Shina Tan~\cite{tan08a}.
Here $A$ is a constant and $a$ is the \textit{s}-wave scattering length.
This is a  wave function that has a divergent
behavior when $r_{i,i'}$ goes to zero, but is square integrable.
In this work, we concentrate on the
$a\rightarrow-\infty$ case,
but the calculation for other values is a straightforward extension.

To satisfy the boundary condition,
and approximately satisfy
the two-body Schr\"odinger equation when unlike pairs are close
together, we choose the
unlike spin Jastrow factor to be
\begin{equation}
f_{\uparrow\downarrow}(r) =
\left \{
\begin{array}{cc}
\frac{d}{r}\frac{\cosh(a_kr)}{\cosh(a_k d)}  & r < D\\
1 & r >D\\
\end{array}
\right . \,,
\end{equation}
and $a_k$ is chosen to make the derivative continuous at $r=D$.
Physically, we expect the healing distance to be of order the
interparticle spacing as born out by the calculations.	The best value
of $D$ is obtained optimizing the variational energy of the system.

Because of the Pauli exclusion principle, the like spin
particles are kept apart. Although a small improvement in variational
values are possible by including a like spin Jastrow factor, we
have chosen to take it to be unity here. It affects neither
the nodal structure nor the boundary condition of the zero-range
interaction, its optimization will be left for future work.

The variational energy of the system, which is
the expectation value of the Hamiltonian is
\begin{equation}
E = \int p(R) E_L(R)dR\;\;, \label{eq:energy}
\end{equation}
where $E_L= \frac{H\Psi_T (R)}{\Psi_T (R)}$
is the local energy and
$p(R) = \frac{\vert \Psi_T(R)\vert^2}{\int dR \vert \Psi_T(R)\vert^2}$
is interpreted as the probability density for a given configuration
$R$.
In a standard variational Monte Carlo, this probability density is sampled
using the Metropolis et al. algorithm~\cite{met53,cep77}.
For $N$ samples, the
variational energy of the system is calculated from
\begin{equation}
\label{eq:var_energy}
E = \frac{1}{M}
\sum_{i=1}^M E_L(R_i)\;\;.
\end{equation}
Because of the divergent behavior of our trial function, the variance
of the $E_L(R)$ diverges when this standard sampling is used. In the
next section we will describe our modifications to control the variance.

In diffusion Monte Carlo~\cite{fou01},
the trial wave function is evolved in imaginary time.
The imaginary
time-dependent Schr\"odinger equation for the system is
\begin{equation}
 -\frac{\partial \psi(R,t)}{\partial t}=(H-\tilde E)\psi(R,t) \,,
\end{equation}
and the solution is
\begin{equation}
\psi(R,t) = e^{-H t}\psi(R) \rightarrow
e^{\left( E_0 - \tilde E \right) t} \psi_0(R) \,,
\end{equation}
where $\tilde E$ controls the normalization of
the wave function in the limit of a long imaginary time.

The energy expectation value as a function of imaginary time
is~\cite{kal74,fou01,cha04}.
\begin{eqnarray}
\label{eq:mixed}
E(t) &=& \frac{\langle \Psi(\frac{t}{2})|H|\Psi(\frac{t}{2})\rangle}
{\langle \Psi(\frac{t}{2})|\Psi(\frac{t}{2})\rangle}
=
\frac{\langle \psi_T|e^{-H\frac{t}{2}} H
e^{-H\frac{t}{2}}|\psi_T\rangle}
{\langle \psi_T|e^{-H\frac{t}{2}}
e^{-H\frac{t}{2}}|\psi_T\rangle}
=
\frac{\langle \psi_T|H e^{-Ht}|\psi_T\rangle}
{\langle \psi_T|e^{-Ht}|\psi_T\rangle}
\nonumber\\
&=&\int dR E_L(R) P(R,t) 
\end{eqnarray}
with the probability density
\begin{equation}
P(R,t) = \frac{\Psi_T(R)\Psi(R,t)}
{\int dR \Psi_T(R)\Psi(R,t)} \;.
\end{equation}

Including the $\Psi_T(R)$ importance function allows us to sample
$P(R,t)$.
The propagation equation in imaginary time is
\begin{equation}
\Psi_T(R)\psi(R,t+\Delta t) = \int dR'\frac{\Psi_T(R)}{\Psi_T(R')}
G(R,R',\Delta t) \Psi_T(R')\psi(R',t) \label{eq:psit}
\end{equation}
where the new walkers can be sampled from
$\frac{\Psi_T(R)}{\Psi_T(R')}G(R,R',\Delta t)$.

Standard diffusion Monte Carlo uses a Trotter breakup to sample
$G(R,R',\Delta t)$. As we noted above, this will fail for the zero-range interaction because its strength goes to negative infinity.
Instead we use the pair product approximation which is commonly
used for path integral calculations\cite{cep95}.
That is we solve for the two-body propagator $g_2(\vec r'_1,\vec r'_2
,\vec r_1, \vec r_2, \Delta t)$ which is the solution of
\begin{eqnarray}
h_2 &=& \frac{p_1^2}{2m}+\frac{p_2^2}{2m} + v(|\vec r_1-\vec r_2|)
\nonumber\\
g_2(\vec r'_1,\vec r'_2,\vec r_1, \vec r_2, \Delta t) &=&
\langle \vec r'_1,\vec r'_2|e^{-h_2 \Delta t}|\vec r_1, \vec r_2\rangle
\,,
\end{eqnarray}
and construct the many-body pair product propagator from
\begin{equation}
\label{eq.pp}
G(R',R,\Delta t) =
\frac{\prod_{ij'} g_2(\vec r'_i,\vec r'_{j'},\vec r_i,\vec r_{j'})}
{\prod_{ij'} g^0_2(\vec r'_i,\vec r'_{j'},\vec r_i,\vec r_{j'})}
G^0(R,R',\Delta t)
\end{equation}
where the zero superscript indicates the solution with zero potential.

The 2-body propagator separates into relative and center of mass
coordinates in the usual way. The center of mass propagates like
a free particle with solution
\begin{eqnarray}
\vec r_{cm} &=& \frac{\vec r_1+\vec r_2}{2}
\nonumber\\
g_{2\ cm}(\vec r'_{cm},\vec r_{cm},\Delta t) &=&
\left( \frac{m}{\hbar^2\pi\Delta t}\right)^{3/2}
e^{-\frac{m \vert \mathbf{r}_{cm}-\mathbf{r}_{cm}' \vert^2}
{\hbar^2\Delta t}}
\end{eqnarray}

The relative coordinates can be
separated into the different angular momentum partial waves, and
in the limit of zero-range, only the s-wave component is different
from the zero potential, free particle result. For the unitary limit,
the result is
\begin{eqnarray}
\label{eq.g2rel}
\vec r_{12}&=&\vec r_1 -\vec r_2
\nonumber\\
g_{2\ rel}( \vec r'_{12},\vec r_{12},\Delta t)
&=&
\left( \frac{m}{4\hbar^2\pi\Delta t}\right)^{3/2}
e^{-\frac{m \vert \vec r'_{12}-\vec r_{12} \vert^2}{\hbar^2\Delta t}}
 + \frac{1}{4\pi^2 r_{12}r_{12}'}
\sqrt{\frac{m\pi}{\hbar^2\Delta t}}
e^{-\frac{m \left(r_{12}+r'_{12}\right)^2}{4\hbar^2\Delta t}}\,.
\nonumber\\
\end{eqnarray}
The first term is the usual free-particle pair propagator, the second
is the additional s-wave amplitude from the interaction. Besides
the usual gaussian, it has
a $\frac{1}{r_{12}r'_{12}}$ prefactor. The importance sampling with a 
trial function with the correct boundary conditions,
multiplies this by $\frac{r_{12}}{r'_{12}}$, this becomes ${r'_{12}}^{-2}$
and this cancels the volume element factor ${r'_{12}}^2$ when
sampled in spherical coordinates. Therefore sampling the two-body propagator,
including importance sampling is straightforward.

The complete two body propagator is
\begin{equation}
g_2(\vec r'_1,\vec r'_2,\vec r_1,\vec r_2,\Delta t)
=
g_{2\ rel}( \vec r'_{12},\vec r_{12},\Delta t)
g_{2\ cm}(\vec r'_{cm},\vec r_{cm},\Delta t) \,.
\end{equation}

\section{Computational details}
\label{sect:details}

We use periodic boundary conditions in a cubic box of side $L$.
The corresponding wave vectors are
particle momentum states are written as plane waves with momentum
$\vec k_i = \frac{2\pi}{L} \left [ n_x \hat x + n_y \hat y +n_z\hat z\right ]$.
We have closed shells for the noninteracting system with
numbers of particles $N = 14,38$, $54$, $66$ and $114$, but shell
effects are essentially nonexistent at the unitary limit.\cite{car03}

We use the standard
Metropolis et al. algorithm
that is used to sample the configurations with probability
density described in \eq{energy}. The maximum displacement of each
particles is adjusted to minimize the autocorrelation.

With the correct
boundary condition,
the delta function in $\nabla^2 f_{\uparrow\downarrow}(r)$
at the origin exactly cancels the potential contribution. However,
since the $f_{\uparrow\downarrow}(r) \sim r^{-1}$ at the origin,
its form looks
like the potential of a point charge at the origin. Its gradient will
look like the electric field of a point charge, and the terms in
the kinetic energy like
$\vec \nabla_i f_{\uparrow\downarrow}(r_{ij'})
\cdot \vec \nabla_i f_{\uparrow\downarrow}(r_{ik'})$  will diverge
when the distance between particles $i$ and $j'$ goes to zero.
However, when this distance goes to zero, all orientations of the
$i-j'$ pair are equally probable. Integrating over this orientation
shows that the divergent terms give zero contribution as the
separation goes to zero. However, the variance from a standard
Metropolis et al. sampling will diverge (as also seen in previous work
when the range of the finite range potential is reduced). We therefore
do not calculate the energy with the standard Metropolis samples. Instead,
before calculating the energy, we make an additional
Metropolis trial move where
we interchange the positions of the closest unlike spin pair. We
use the heat-bath acceptance probability for this move
\begin{equation}
A(R\rightarrow R') = \frac{\Psi_T^2(R')}{\Psi_T^2(R')+\Psi_T^2(R)}
\end{equation}
and calculate the energy using the method of expected values
\begin{equation}
E_L(R) = A(R\rightarrow R')E_L(R')
+[1-A(R\rightarrow R')]E_L(R) \,.
\end{equation}
The diverging part of the variance cancels in the limit of the pair
distance becoming small. We have found it adequate to interchange only
the closest pair for the system sizes we have used. The method can
be made extensive while maintaining a polynomial complexity
by calculating the single particle part of the
energy and including more such exchanges if needed.

After the variational calculation, the trial wave function is evolved in
imaginary time in accordance with \eq{psit}. In previous diffusion
Monte Carlo calculations using the pair propagator\cite{pudliner97}
the two or more points of the
free-particle propagator $G^0$ were sampled and the pair product
used to sample among these. Here, because the physics is dominated
by the pair interaction, we first sample what we call
the independent pair propagator. That is we find the closest unlike
spin pair and include it. We then find the next closest pair that
does not contain either of the closest pair particles. We continue
in this way until all the particles are paired. Taking only
these terms in the products in Eq. \ref{eq.pp} gives us the
independent pair propagator which we call $G_{ip}$.

We sample the independent pair propagator with approximate
importance sampling. We introduce a cut off distance which is
a few times the width of the free particle gaussian, corresponding
to the distance where the second interaction term in the two-body relative
propagator is negligible. For pairs beyond this, their propagation
is not affected by the interaction, and we can sample just the
first term of Eq. \ref{eq.g2rel}. For pairs within this distance,
we sample the center of mass from its gaussian, and we sample
the two-body relative propagator including the $r^{-1}$ term from the
importance function as mentioned above. This results in our sampling
$R_1$ from
\begin{equation}
P_1(\vec R_1,R) = \frac{\Psi_I(R_1)}{\Psi_I(R)}G_{ip}(R_1,R)
\end{equation}
where
$\Psi_I(R_1)=\prod_{ij'}\frac{1}{r_{ij'}}$
is the product of the relative distances for
all the independent pairs within the cut off distance.

Just as in the variational calculation, we must cancel
the local energy divergences.
Our modification of the sampling is based on the method commonly
used for quantum Monte Carlo calculations in nuclear physics\cite{pudliner97}.
There, the free many-body gaussian propagator is sampled. Since changing
the sign of all the gaussian samples gives an equally good sample,
the importance sampling is included by choosing between these according
to the relative values of the importance sampled propagator.
Adapting a combination of this nuclear physics method
and the method we used for the variational calculation,
we determine the
closest pair and choose four samples given by $R_1$, the original sample,
$R_2 = {\cal N}(R_1)$, $R_3 = {\cal I}(R_1)$, $R_4={\cal I}(R_3) =
{\cal I}({\cal N}(R_1))$, where the operator ${\cal N}$ changes the
sign of the Gaussian sample $\Delta R_G \rightarrow -\Delta R_G$, and
the ${\cal I}$ operator interchanges the positions of the particles
in the pair with the closest separation distance.  The probability of
choosing a particular configuration $R'$ is given by the sum over the
four probabilities associated with the samples
\begin{eqnarray}
P(R',R) &=&
\frac{\frac{\Psi_T(R')}{\Psi_T(R)}G(R',R)}
{\sum_j\frac{\Psi_T(R_j)}{\Psi_T(R)}G(R_j,R)}
\left [ \sum_{i=1}^{4} P_1(R_i,R)\right ] \,,
\end{eqnarray}
where $P_1(R_3,R) = P_1(R_4,R) = 0$ if the relative coordinate of the
closest pair is less than a input parameter $r_{int}$. Consequently,
the weight for choosing the particular configuration is
\begin{eqnarray}
W(R') &=& \frac{\sum_j\frac{\Psi_T(R_j)}{\Psi_T(R)}G(R_j,R)}{ \sum_j
\frac{\Psi_I(R_j)}{\Psi_I(R)} G_{ip}(R_j,R)}\;\;.
\end{eqnarray}
Just as in the variational calculations, the divergence in the
local energy is canceled, and the variance is well controlled.

\section{Results and Discussion}
\label{sect:results}

\begin{figure}
\begin{center}
\includegraphics[width=0.9\linewidth]{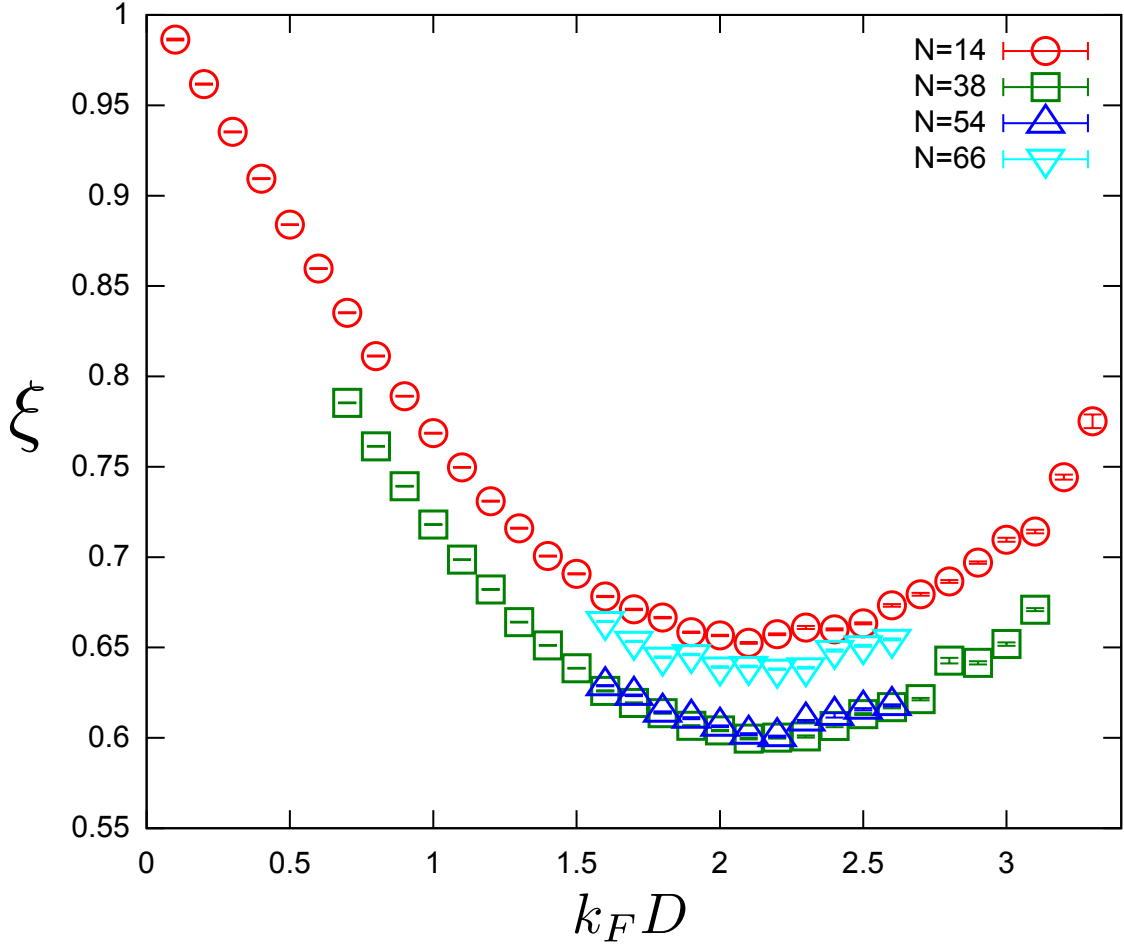}
\caption{\label{Fig:ener_rang}(Color online) Ratio between the variational
energy per particle and the noninteracting Fermi gas energy as a function
of the Jastrow term range. Results for $N=14$ ({\color{red}$\circ$}),
$N=38$ ({\color{green}$\square$}), $N=54$ ({\color{blue}$\triangle$}),
$N=66$ ({\color{cyan}$\triangledown$}) particles.
}
\end{center}
\end{figure}

\begin{figure}
\begin{center}
\includegraphics[width=0.9\linewidth]{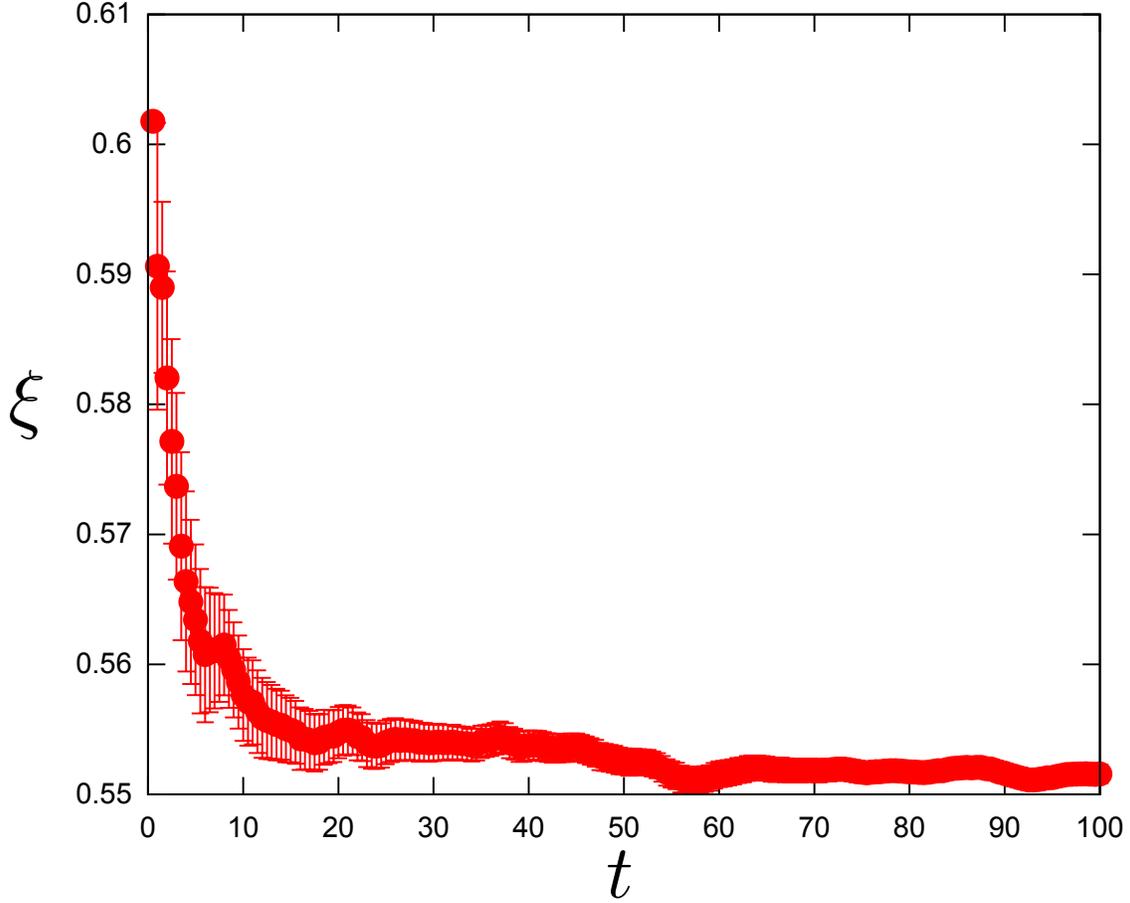}
\caption{\label{Fig:ener_time}(Color online) Ratio between the energy
per particle and the noninteracting Fermi gas energy as a function of the
imaginary time in the DMC calculation. Results for $N=14$ particles. Each
time step $\Delta t = 10^{-3}\hbar/E_{FG}$.
}
\end{center}
\end{figure}

We performed a large number of computational simulations with different values
of variational parameters to minimize the energy of the system. The
\fig{ener_rang} shows the variational energy for different number of particles
as a function of the healing distance. The minimal variational
energy for all the cases is obtained for the same value $k_F D=2.1$. This value
is used for the different numbers of particles in this paper. The optimized
variational energy per particle is $E/N = 0.653(1) E_{FG}$ estimated for the
system containing $14$ particles. The \fig{ener_time} shows the ground state of
the system calculated through the mixed energy estimator of \eq{mixed} as a
function of the imaginary time. Each time step is $\Delta t = 10^{-3}\hbar/E_{FG}$ and the results show that the asymptotic value
$t\rightarrow \infty$
is obtained after
about $t = 20\hbar/E_{FG}$. The DMC energy for the system with 14 particles is
$E/N = 0.551(1)$, significantly below the variational energy result.

\begin{figure}
\begin{center}
\includegraphics[width=0.9\linewidth]{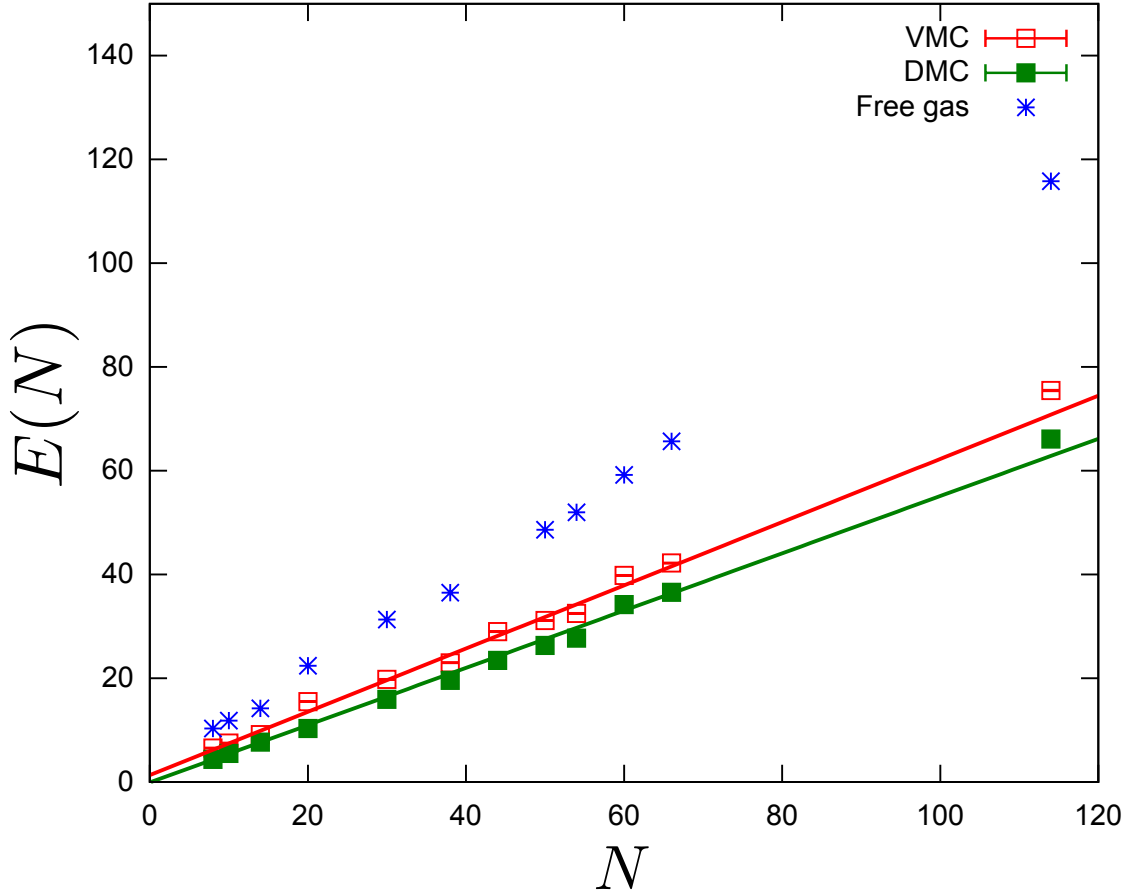}
\caption{\label{Fig:energy}(Color online) The energy of the system as a
function of the number of particles in the box. The red empty and green
filled squares are the VMC and DMC energy results, respectively. Results
for the non-interacting free gas are displayed for a comparison
({\color{blue} $\ast$}). Energy in units of $E_{FG}$.
}
\end{center}
\end{figure}

The \fig{energy} shows the ground state energies of the system
as a function of the number of particles in the simulation.
The red squares represent
the variational results and the fitted straight red line results $E =
0.61(2)E_{FG}$. The variational energy upper bound obtained by Chang and
coworkers~\cite{cha04} is $E = 0.62(1)E_{FG}$ that is in good agreement with our
result for the energy as expected for the similar forms for the trial
wave functions. The main difference is our results do not need
to be extrapolated to zero-range.
The diffusion Monte Carlo energy is, of course, much lower.
The
green filled squares in the \fig{energy} show the energies for the
diffusion Monte Carlo
calculation evolving the Jastrow-Slater wave function in imaginary time. The
fitted green straight line gives $E = 0.55(1)E_{FG}$ in very good
agreement with the available result from the literature~\cite{gio08} $E =
0.56(1)E_{FG}$ for the Slater-Jastrow wave function.

\begin{figure}
\begin{center}
\includegraphics[width=0.9\linewidth]{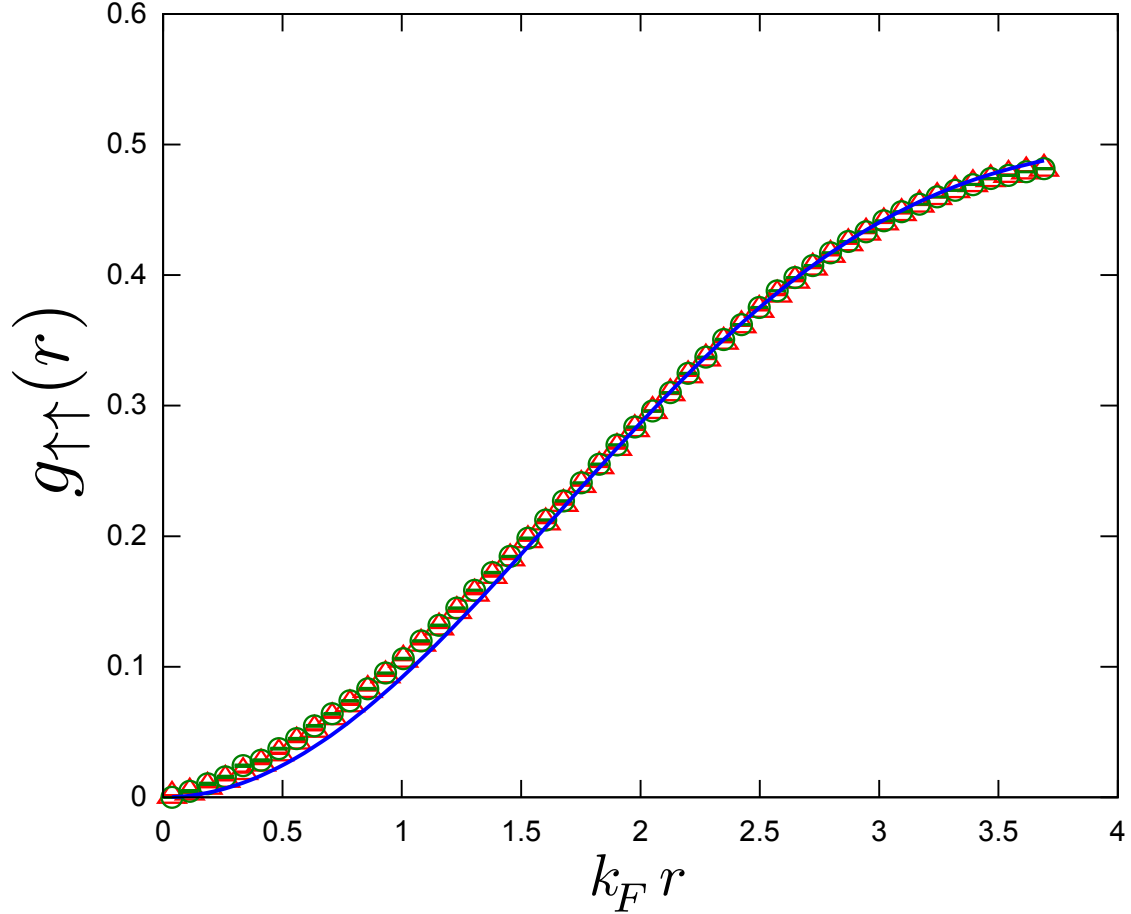}
\caption{\label{Fig:gr_like}(Color online) Pair correlation function of
like spin pairs calculated for $N=14$ particles. VMC and DMC results are
represented by the green circles and the red triangles, respectively. The
solid line shows the behaviour of the pair correlation function of the
non-interacting Fermi gas \eq{free}.
}
\end{center}
\end{figure}

We have calculated the
spin dependent pair correlation function.
\fig{gr_like} shows the results for
$g_{\uparrow\uparrow}(r)$ with a comparison with the non-interacting
gas. For like spin pairs, the pair correlation function goes to zero
at short distances because of the Pauli principle. The result for the
non-interacting Fermi gas~\cite{ast04} is described by
\begin{equation}
g_{\uparrow\uparrow}(r)=g_{\downarrow\downarrow}(r) = 1 - \frac{9}{(k_F r)^4}\left( \frac{\sin(k_F r)}{k_F r} - \cos(k_F r)\right)^2\;\;. \label{eq:free}
\end{equation}
In the \fig{gr_like} one can see reasonable agreement between our results and
the pair correlation function of the non-interacting Fermi gas, however
the interaction modifies this and the result will be changed using
diffusion Monte Carlo.
As expected, $g_{\uparrow\downarrow}(r)$ has a
$1/(k_Fr)^2$ divergent behaviour when $r$ goes to zero~\cite{gio08}.
Both 
$g_{\uparrow\downarrow}(r)$ and
$g_{\uparrow\uparrow}(r)$ are normalized to go to $\frac{1}{2}$ at
large distances.

\begin{figure}
\begin{center}
\includegraphics[width=0.9\linewidth]{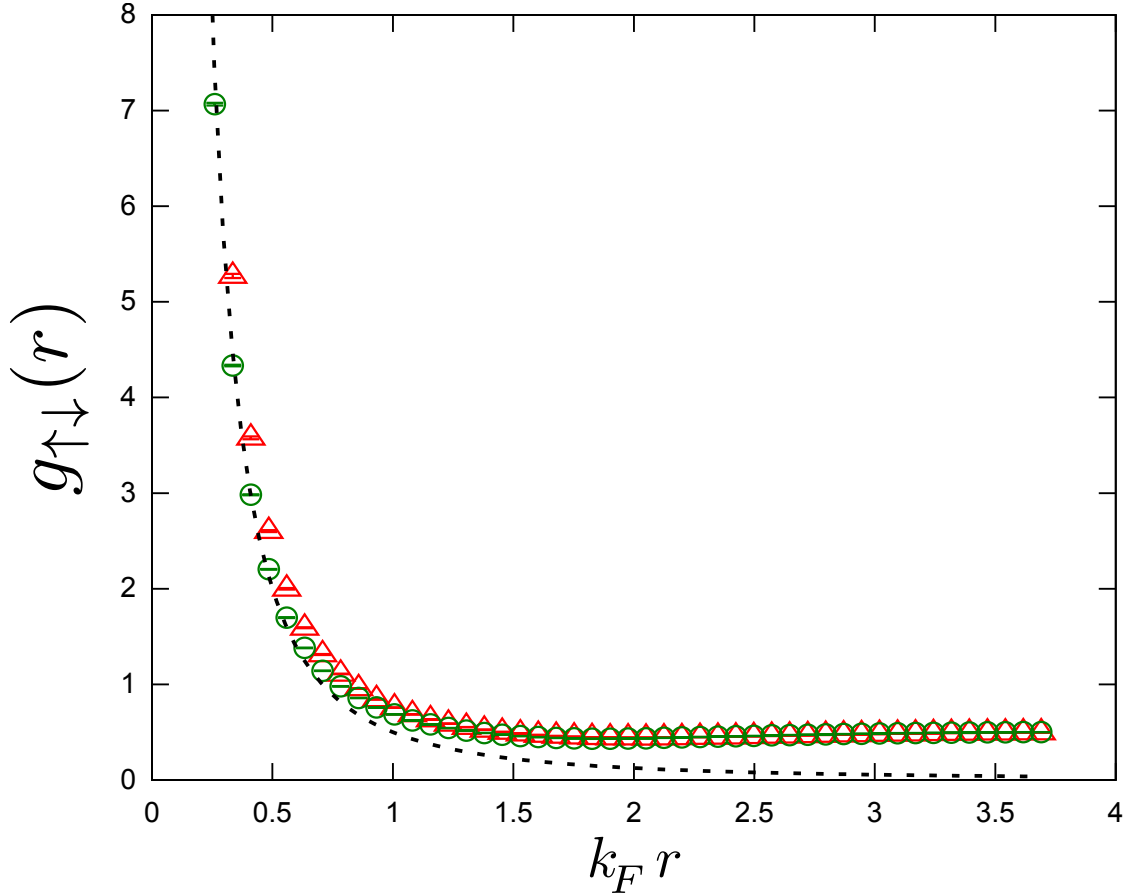}
\caption{\label{Fig:gr_unlike}(Color online) Pair correlation function
of unlike spin pairs for the system containing $N=14$ particles. VMC and
DMC results are represented by the green circles and the red triangles,
respectively. The dashed line is the function $1/(k_Fr)^2$.
}
\end{center}
\end{figure}

\section{Summary}
\label{sect:summ}

In this work we have demonstrated that a system formed by
ultracold fermions
at the unitary regime can be studied using 
a zero-range interaction.
In our Monte Carlo calculations
we have devides variational trial wave functions
with the correct boundary conditions, and we have developed improved
sampling techniques to give a controlled variance in this zero-range
limit.

The variational and diffusion Monte Carlo
results are in good agreement with those reported in the
literature for extrapolations of short but finite ranged potentials.
We expect that our methods would could also be applied
to improve the results with finite range potentials.

Further work optimizing the
BCS wave functions can improve significantly the results reported in this
paper. Automatic optimization of the BCS wave function for zero-range
interactions as well as calculating other observables
in ultracold Fermi gases is an ongoing.

\section*{Acknowledgments}
R.P. was financially supported by the Brazilian agency
CAPES (Coordena\c{c}\~ao de Aperfei\c{c}oamento de Pessoal de N\'{i}vel
Superior) Proc. n. 11540/13-3. Part of the computations were performed
at the CENAPAD high-performance computing facility at Universidade
Estadual de Campinas and at the laboratory of scientific computation of
Universidade Federal de Goi\'as. SAV acknowledges the financial support from
FAPESP under grant 2010/10072-0. KES would like to thank
the Instituto de F\'{\i}sica Gleb Wataghin,
Universidade Estadual de Campinas - UNICAMP, for their hospitality
where some of this work was performed. KES was partially
supported by CAPES-PVE-087/2012
and by the National Science
Foundation grant, PHY-1404405.

\bibliographystyle{apsrev}

\end{document}